\documentclass[english,prl,superscriptaddress,preprintnumbers,twocolumn
]{revtex4} 
\usepackage{graphicx,  bm, color, babel}

\makeatletter

\def\be{\begin{equation}}
\def\ee{\end{equation}}
\def\bea{\begin{eqnarray}}
\def\eea{\end{eqnarray}}

\begin{document}

\title{The isotropic blackbody CMB as evidence for a homogeneous universe}

\author{Timothy Clifton}
\affiliation{School of Physics and Astonomy, Queen Mary University of London, UK.} 

\author{Chris Clarkson}
\affiliation{Department of Mathematics and Applied Mathematics, University of
  Cape Town, South Africa.}

\author{Philip Bull}
\affiliation{Department of Astrophysics, University of Oxford, UK.}

\begin{abstract}
\vspace{5pt}
The question of whether the Universe is spatially homogeneous and isotropic on
the largest scales is of fundamental importance to cosmology, but has
not yet been answered decisively.  Surprisingly, neither an isotropic
primary CMB nor combined
observations of luminosity distances and galaxy number counts are
sufficient to establish such a result.  The inclusion of
the Sunyaev-Zel'dovich effect in CMB observations, however,
dramatically improves this situation. We show that even a solitary
observer who sees an isotropic blackbody CMB can conclude that the
universe is homogeneous and isotropic in their causal past when the
Sunyaev-Zel'dovich effect is present.  Critically, however, the CMB
must either be viewed for an extended period of time, or CMB photons
that have scattered more than once must be detected.  This result
provides a theoretical underpinning for testing the Cosmological
Principle with observations of the CMB alone.
\vspace{5pt}
\end{abstract}

\maketitle

The current concordance model of cosmology is based on the homogeneous and 
isotropic Friedmann-Lema\^{\i}tre-Robertson-Walker (FLRW) solutions of
Einstein's equations.  The high degree of symmetry assumed in 
these solutions makes them sufficient to explain the near perfect isotropy of the Cosmic Microwave Background (CMB) 
and other astrophysical observables, but it remains to be demonstrated whether 
or not they are the {\it only} spacetime geometries that are
compatible with the data.  This question is  
particularly pertinent due to the apparent necessity that more than 
$95\%$ of the matter content of the Universe must be in the form of dark 
energy and dark matter in order for the concordance model to be made compatible
with observations. The inferred existence of these substances holds such
profound consequences for our understanding of basic physics that
establishing the validity of the assumed FLRW geometry is now an
imperative. So, what observables are required in order to prove the universe is FLRW on large scales?

An important step toward answering this question was provided by Ehlers, Geren and Sachs (EGS)~\cite{EGS}, and later fleshed out by others~\cite{FerMorPor99,Clarkson:1999yj} (see~\cite{2010CQGra..27l4008C} for a review). These authors used the Copernican Principle, that we are typical observers, to show that isotropy of the CMB about every point in a region of spacetime is only possible if the geometry of spacetime in that region is spatially homogeneous and isotropic.
This result is perturbatively stable in the sense that {\it near} isotropy of the CMB implies {\it near} homogeneity of spacetime, although this  requires
extra assumptions about unobservable
quantities~\cite{1995PhRvD..51.1525M,1995PhRvD..51.5942M,1995ApJ...443....1S,Rasanen:2009mg}. An alternative proof of spatial homogeneity using luminosity distances, that also relies on the 
Copernican Principle, was found by Hasse and Perlick~\cite{HP}. 
While compelling these theorems all require observations to be made at {\it all} 
points in a region of spacetime to make definite
conclusions. Isotropy of the CMB on our own sky is not even sufficient
to determine that our local region of 
space is isotropic around us \cite{Roy}.

Alternatively, the authors of~\cite{MM} have shown that in order to determine
whether the  Universe is isotropic around us it is necessary and
sufficient to have isotropic observations of luminosity distances,
number counts, lensing, and angular peculiar velocities at every
redshift, and in every direction.  To then determine spatial homogeneity
requires an extra independent observable beyond these four, unless one is
prepared to specify the value of $\Lambda$ {\it a priori} (assuming
dark energy is due to the cosmological constant \footnote{If
dark energy is a dynamical field then we actually have an extra
functional degree of freedom to determine observationally,
complicating the situation even more.}) \cite{Hellaby}.
While this prescription for determining spatial homogeneity and isotropy has 
the important quality of relying solely on directly observable quantities, rather
than the Copernican Principle, it also requires large amounts
of information from a number of different observables. 

Here we show that inclusion of the the Sunyaev-Zel'dovich (SZ) effect
when considering CMB observations allows one to retain the minimal
observational  requirements of EGS, while removing the assumption of
the Copernican  Principle. The SZ effect is due to the scattering of
CMB photons by charged  
matter, and has already been shown to be a powerful tool for constraining radial
inhomogeneity within the class of cosmological models constructed from
the Lema\^{i}tre-Tolman-Bondi solutions
\cite{2008JCAP...09..016G, 2010CQGra..27f5002G, 2011PhRvL.107d1301Z,
  2011CQGra..28p4005Z, 2011arXiv1108.2222B, 2010JCAP...10..011Y,
  2011CQGra..28p4004M, 2011JCAP...02..013C, Goodman,
  CaldwellStebbins}. 
We extend these previous studies to consider the potential of the SZ
effect to act as proof of FLRW geometry, rather than simply as a tool for
constraining particular deviations away from it. 
This results in a stronger 
statement than that of EGS, as it requires observations made by only a single 
observer, rather than from all observers in a region of spacetime. It is 
a much less demanding statement than the result of~\cite{MM}, 
as it requires observations of the CMB only (although the CMB must
be viewed for an extended period of time, or photons that have
scattered more than once must be detectable).

The SZ effect is often divided into two different contributions; the thermal SZ effect (tSZ)
\cite{tSZ} and the kinematic SZ effect (kSZ) \cite{kSZ}.
The former of these describes the transfer of energy from the hot
electrons in the intra-cluster medium to the cooler photons of the CMB.
This is the easiest of the two effects to detect observationally, but
is the least important for our current considerations.  We will assume
here that the tSZ effect is well understood, and can be removed from the CMB 
signal along with other unwanted foreground sources (a process that will certainly be complicated by relativistic corrections).

\begin{figure}[t]
\includegraphics[width=1.05\columnwidth]{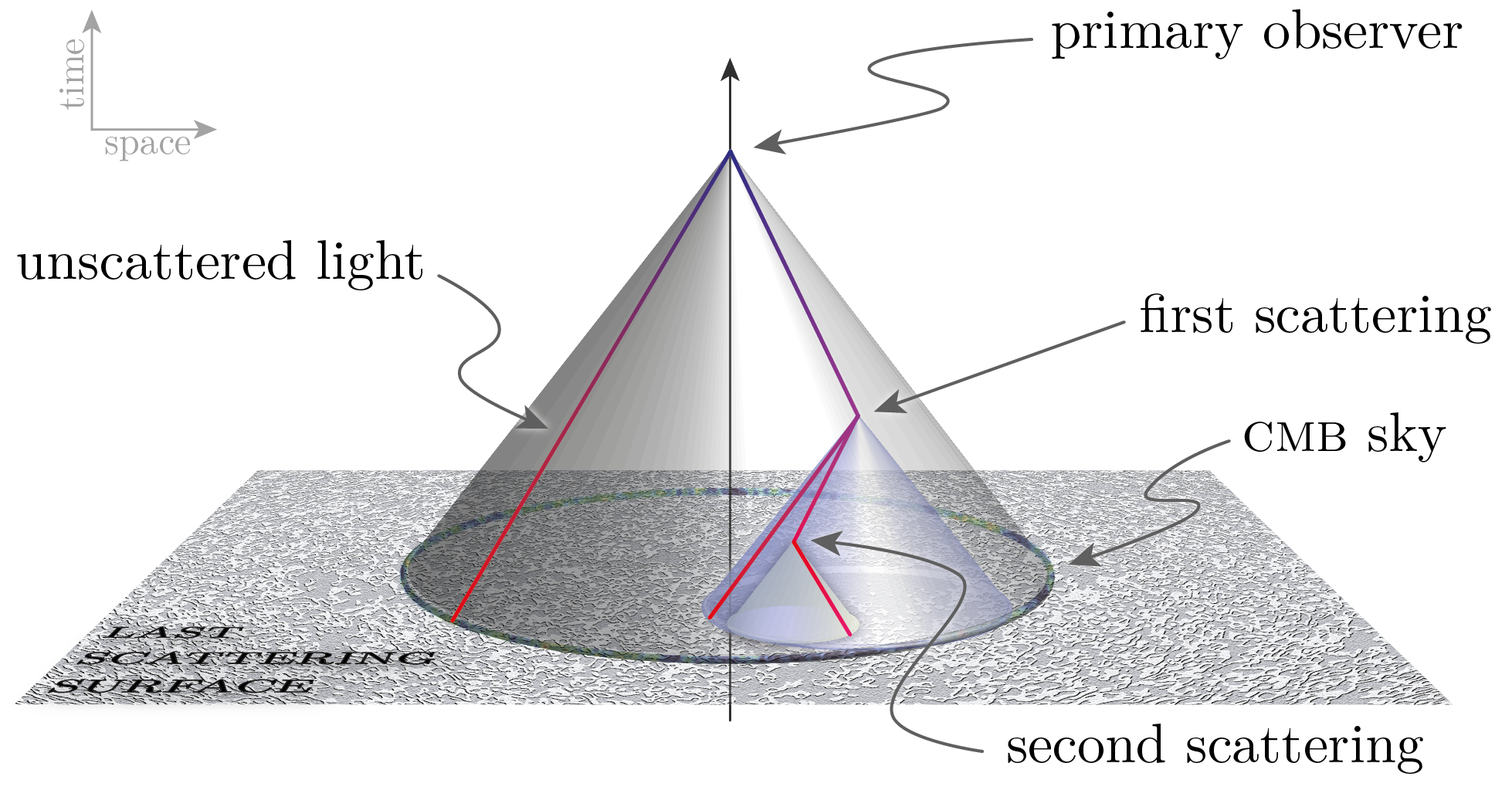}
\vspace{-0.2cm}
\caption{The kSZ effect provides information about 
  the CMB sky at other points on our past lightcone. 
}
\label{fig}
\vspace{-0.35cm}
\end{figure}

The kSZ effect also alters the spectrum 
of the scattered light, but this is due to the anisotropy seen in the CMB sky 
of the scatterer rather than any transfer of energy from baryons to photons in 
the baryon rest frame.  In an FLRW spacetime, any such anisotropy is due to 
the peculiar motion of the scatterer. In the rest frame of the scatterer, the 
re-scattered light then maintains the same distribution function it had before the 
scattering event (all other changes being encapsulated in the tSZ), so 
that an observer in the rest frame of the CMB must see radiation that undergoes 
a Lorentz boost after scattering. For the case of blackbody radiation this 
corresponds solely to a change in temperature of the scattered radiation. This 
mechanism therefore provides, in effect, a set of mirrors that allows us to view the 
CMB from different locations.  We shall therefore refer to the light scattered into 
our line of sight as being {\it reflected} by the scatterer (which we will 
refer to as a cluster, for simplicity).

The picture described above is valid in an FLRW universe with an isotropic 
radiation field, but in the present study this is exactly the thing we 
want to prove the existence of. It is therefore necessary to generalise 
the existing concept of the kSZ effect.

The picture we have for this generalised scenario is illustrated in
Fig.~\ref{fig}.  As we look back along our past null cone we will see
the reflecting clusters, whose own past null cones coincide with ours 
in one direction, but otherwise crosses the last scattering surface
within our causal past.  If there exist sufficiently many clusters,
we will receive photons from every part of the last scattering
surface that we are causally connected to, rather than from just the
single sphere that we observe directly.  We assume that the formation of 
the last scattering surface proceeds in thermal equilibrium, so that the
emitted radiation is blackbody, and that the Universe is optically thin at 
all times and everywhere after last scattering. Scattering off the clusters 
can then result in a possible temperature change, but the spectrum must remain
a blackbody as Lorentz transformations at the point of
reflection, and cosmological evolution in an arbitrary spacetime, both
preserve the form of a blackbody spectrum.  Also illustrated in
Fig.~\ref{fig} is the possibility of photons being scattered off two
clusters before they reach us, which we will return to later.

Let us denote the incident temperature in each direction on the reflecting
cluster's sky as $T_i=T_i(\theta,\phi,z)$, where $\theta$ and $\phi$ are
spherical polar coordinates on their sky, chosen  such that $\theta
=\pi$ is the direction of the eventual observer (us), and $z$ is the
redshift of the cluster on the eventual observer's sky.  The occupation
number of photons received from a particular direction $(\theta,
\phi)$ on the cluster's sky can be written as $N_i=\mathcal{B}(\nu,T_i)$, where
$
\mathcal{B}(\nu,T) = (e^{{\nu}/{T}}-1)^{-1},
$
are the occupation numbers of a blackbody spectrum with frequency
$\nu$, and where we have set $k_B=h=1$.   
The fraction of light that is reflected towards the observer from
every direction on the cluster's sky is given by the Thomson
cross-section, which,
after the effects of the tSZ effect have been removed, gives the
occupation number 
of the reflected light in the rest frame of a particular cluster as
$
N_r(\nu,T_i,z) = \frac{3}{16\pi} \int \tau (1+\cos^2
\theta) \mathcal{B} (\nu,T_i) \sin \theta  d \theta d \phi
$, where $\tau\ll 1$ is the electron-scattering optical depth of the
cluster, which is assumed to fill the telescope beam.
We now want to know the conditions on incident radiation on the cluster, 
$T_i=T_i(\theta, \phi,z)$, for the sum of the reflected light and the 
unscattered light to have a blackbody spectrum when it is observed 
at $z=0$. Recall that blackbody spectra are unchanged after propagating though
spacetimes with arbitrary curvature, up to a change in temperature by
one factor of redshift \cite{Ellis}, so we can write the observed temperature 
of any blackbody distribution as $\bar{T}=T/(1+z)$ (where $z$ is the redshift 
at which it had temperature $T$). For a continuous distribution of matter we 
can also write the reflected radiation in some interval $\Delta z$ along one of 
our own past-directed null geodesics as $N_r(\nu, T_i, z) \Delta z$.
The distribution function of photons that 
make it to us is then given by
$
N_{\rm tot}= \mathcal{B}(\nu, \bar{T}_c) + \int N_r(\nu, \bar{T}_i,z)
dz - \int N_r(\nu, \bar{T}_c,z) dz, 
\nonumber
$
where $T_c(z)=T_i(0,0,z)$ is the temperature
of the unscattered light at redshift $z$ \footnote{The integrals in this 
expression are understood to be along the past nullcone, which may not be 
monotonic in $z$.}.  The first term on the
RHS of this equation is the contribution from the unscattered CMB, the second 
term is CMB radiation that is scattered towards us (that we would otherwise 
not be able to observe), and the third term is the CMB radiation that is
scattered away from us (that would otherwise reach us in the absence of any 
scattering).  For $N_{\rm tot}$ to be a blackbody with some
temperature $T_0$ we then require $N_{\rm tot}=\mathcal{B}(\nu,T_0)$.
If we change variables to $x=-\cos \theta -\frac{1}{3} \cos^3 \theta$,
and expand each term in a power series around its temperature, we get
\bea
\hspace{-2cm}
&&\sum_{n,k}  c_n {n \choose k} (-1)^k \nu^{n-k}  \Bigg[ \left(
 \bar{T}_c^{k-n} - T_0^{k-n}
\right) \nonumber \\ &&\qquad + \int
 \frac{3 \tau}{16 \pi} \left( \bar{T}_i^{k-n} -
  \bar{T}_c^{k-n} \right) dx d\phi dz \Bigg]= 0, \label{eqn_cn}
\eea
where $c_n = (-1)^n e A_n(e)/n!(e - 1)^{n+1}$, and $A_n(x)$ are the Euler 
polynomials.  It can be shown that $A_n(e)$ are positive definite, so that the 
$c_n$ have sign $(-1)^n$.  Now, Eq.~(\ref{eqn_cn}) must be true for each 
value of $n-k$, as nothing here is a function of $\nu$ except $\nu^{n-k}$ 
itself. For $n-k=j$ we therefore have
\bea
\hspace{-2cm}
&&\sum_{n=0}^{\infty}  c_n {n \choose j} (-1)^{(n-j)} \Bigg[ \left(
\bar{T}_c^{-j}-T_0^{-j}
\right) \nonumber \\ &&\qquad+ \int
\frac{3 \tau}{16\pi} \left(
  \bar{T}_i^{-j}-\bar{T}_c^{-j} \right) dx d\phi dz
  \Bigg]=0, \label{eqn-freq-expn} 
\eea
for all $j \neq 0$. In the $j=1$ and $j=2$ cases, we also have
$\sum_{n=0}^{\infty} n c_n (-1)^{n-1} \neq 0$
and
$\sum_{n=0}^{\infty} n (n-1) c_n (-1)^{n-2} \neq 0$,
as $c_n (-1)^{n-1} <0$ and $c_n (-1)^{n-2} > 0$ for all $n$. It must then 
be the case that
\bea
&&
\left[1-\frac{3}{16 \pi} \int \tau dx d\phi dz \right] \left(
\frac{1}{\bar{T}_c}
-\frac{1}{T_0} \right)^2 \nonumber \\&+&
\int  \frac{3 \tau}{16\pi}\left( \frac{1}{\bar{T}_i}
-\frac{1}{T_0} \right)^2 dx d\phi dz=0.
\eea
For $\tau \neq 0$ and $\int \tau dx d\phi dz < 16 \pi/3$ we
therefore have that $\bar{T}_i^{-1}=\bar{T}_c^{-1}=T_0^{-1}$.  The first of
these conditions is that there should exist scatterers everywhere, and the
second is that the amount of reflected radiation must be
less than the amount of the incident radiation (from the Beer-Lambert
law).  We therefore have that $T_i(\theta,\phi,z)=T_c$ for every $\theta$
and $\phi$, at every $z$ where $\tau \neq 0$. We also have that
$T_c=(1+z) T_0$, so that the temperature of the observed CMB must be
the same as the emitted CMB, up to a factor of redshift. This result shows
that the CMB must be isotropic about every reflecting cluster, and is 
essentially due to the fact that blackbody spectra of different
temperatures cannot be summed to give another blackbody spectrum~\cite{bb1}.  

This result tells us that if the CMB was emitted from a thermal process as a 
blackbody, and is observed as a blackbody today, then the CMB sky at every 
point on our past lightcone must be isotropic.  Any anisotropies at any point 
on our past light cone would cause distortions in the spectrum of radiation we 
observe. Surprisingly, this is not yet restrictive enough to require
FLRW geometry. 

Up to this point, we have only considered the CMB sky of observers at other points 
on our past null cone.  This is not sufficient to establish either homogeneity 
or isotropy of space around any point, however, as we also require information about 
derivatives of geometric quantities and the matter content of the
Universe in order to propagate information off our past null cone.

The starting point for this is the Boltzmann equation for photons,
which in general involves a collisional term for the Thomson
scattering.  This term is proportional to changes in the
distribution function~\cite{Maartens:1998xg}, and as we have shown
that a vanishing kSZ effect implies isotropy of the CMB about every
cluster, this means that the collision term must vanish at every
cluster where the kSZ effect vanishes.
Hence, it is sufficient for us to consider the collisionless Liouville
equation.  This tells us that if every time-like observer following a
congruence $u^a$ sees an isotropic radiation field then $u^a$ must be
(parallel to) a conformal Killing vector, and the spacetime must be
conformally stationary~\cite{EGS}.  The anisotropic pressure evolution equation
that is derived from the quadrupole of the Liouville equation then
tells us that $u^a$ must be shear-free, and that the acceleration
$A_a=u^b\nabla_b u_{a}$ and expansion rate $H=\frac{1}{3} \nabla_au^a$ must satisfy 
$
\nabla_{[a}\left(A_{b]}-Hu_{b]}\right) = 0\, 
$. One now needs to make assumptions about the matter content in order to
make further statements.  For an irrotational, geodesic perfect
fluid, it follows that the spacetime must be
FLRW~\cite{1997icm..book.....K}. The radiation-only case is the original
EGS result~\cite{EGS}.
In the case of a mixture of dust, radiation and dark energy in the
form of a scalar field (for which $\Lambda$ is a special case) a little more work is
required~\cite{Clarkson:1999yj} because one cannot assume {\it
  ab initio} that the gradient of the scalar field is aligned with the
dust observers, but isotropy implies it must be and so the result still holds.  
The EGS  theorem holds in general scalar-tensor theories of
gravity~\cite{Clarkson:2001qc}.

The final stage of our result therefore requires us to show that the
CMB must be isotropic inside our past lightcone, as well as on it.  We
can see two possible ways of doing this: 

(1). If we observe the CMB for a finite interval of time \cite{extra1}.
This would allow us to receive information about the CMB sky of all observers in the 
4-dimensional region of spacetime swept out by our past null cone over this 
interval. If no kSZ effect is measured at any time, then one can infer that 
the entire region is filled with clusters that see isotropic CMB radiation. 
The region must therefore have FLRW geometry, and taking any surface within it 
as an initial Cauchy surface, we can establish that our entire causal past 
must also be FLRW.

(2). If we can observe CMB radiation that has been scattered more
  than once \cite{extra2,extra3}, as was suggested in the original paper by Sunyaev and
  Zel'dovich \cite{kSZ}.  This situation is illustrated
by the existence of the `second scatterer' in Figure~\ref{fig}.  If
such scattering is observable the CMB sky of the second
scatterers must also be isotropic if we are still to observe the
  re-scattered CMB photons as being a 
blackbody (the proof of this can be found in the Appendix). Note that
only two scatterings are required, as this is sufficient to show that
the CMB must be isotropic around every point in our causal past.

The situation we have considered here is of course highly idealized:
The CMB is neither exactly isotropic nor blackbody, and our 
treatment of the scattering events themselves is also idealized. In reality, 
we see only a limited number of scatterings that must necessarily only happen
at relatively late times (when structure has formed). Also, the removal 
of the tSZ effect will undoubtedly always be imperfect, as will the subtraction of 
other foreground sources. Furthermore, we have been somewhat optimistic in 
considering that it may be possible to observe the CMB for an extended
period of time, or that double scattering events can be detected. Nevertheless, we have demonstrated what is required to show that the 
Universe is FLRW using the CMB alone, without assuming anything about the 
symmetries of spacetime on the largest scales. Our result holds provided 
dark energy can be described as a scalar field, and holds for general 
scalar-tensor theories of gravity too.

To make these ideas more realistic they need to be shown to be
perturbatively stable, which is non-trivial \cite{1995PhRvD..51.5942M,
  1995ApJ...443....1S, 1995PhRvD..51.1525M}.  An application to the real universe will also require careful consideration of the consequences of imperfect observations and noncontinuous scatterers. We leave this for future work.  
\vspace{10pt}

{\bf Appendix: Multiple scatterings.} 
\vspace{10pt}
\newline
Let us denote quantities evaluated at the primary observer (us) with
a subscript $0$, and those evaluated at the first and second
scatterings with $A$ and $B$, respectively.  The redshift of a first scatterer, as
measured by the primary observer, is then $z_A$, and the redshift of
a second scatterer, as measured by a first scatterer, is $z_B$.
Angular coordinates on the sky of the first and second scatterers will
be written as $(\theta_A, \phi_A)$ and $(\theta_B, \phi_B)$. The
temperature of CMB radiation measured on these scatterers' skies are
then $T_A=T_A(\theta_A, \phi_A, z_A)$ and $T_B =
T_B(\theta_A, \phi_A, z_A, \theta_B, \phi_B, z_B)$. Using this
notation, the occupation number of CMB photons at the primary observer is
\bea
\nonumber
&&N_{\rm tot} = \mathcal{B}(\nu, \bar{T}_c) + \frac{3 }{16\pi} \int
\tau(Y_A) \left[ \mathcal{B}(\nu, \bar{T}_A) -\mathcal{B}(\nu, \bar{T}_c)  \right] dY_A\\
&&+ \frac{9  }{(16\pi)^2} \int \tau(Y_A) \tau(Y_B) \left[
\mathcal{B}(\nu,\bar{T}_B) -  \mathcal{B}(\nu, \bar{T}_A)  \right] dY_A dY_B ,
\nonumber
\eea
where we have written $Y_A=\{x_A,\phi_A,z_A\}$, 
$dY_A = dx_A d\phi_A dz_A$, etc. The first term on the RHS
of this equation is from the unscattered CMB, and the second
term is from the CMB light scattered toward and away from
us by photons that are scattered once.  The third term on the RHS is
new, and corresponds to photons 
that are scattered towards and away from us by double scatterings.
The calculation now proceeds as in the single scattering case,
and results in
\bea
&0& =\left[1-\frac{3}{16\pi}\int \tau(Y_A)   dY_A \right] \left ( \frac{1}{\bar{T}_c} -
\frac{1}{T_0} \right )^2 
\nonumber
\\&+&\frac{9  }{(16 \pi)^2}\int \tau(Y_A) \tau(Y_B) \left ( \frac{1}{\bar{T}_B} -
\frac{1}{T_0} \right )^2 dY_A dY_B  \nonumber \\
&+&\frac{3  }{16 \pi} \int \tau(Y_A) \left[1 - \frac{3}{16\pi}\int \tau(Y_B)   dY_B
  \right] \left ( \frac{1}{\bar{T}_A} - \frac{1}{T_0}
\right )^2 dY_A .  \nonumber
\eea
For $\tau \neq 0$ and $\int  \tau dY< 16\pi/3$ we therefore have that
$T_A=T_B=T_c=(1+z) T_0$.  
The inclusion of third and higher order scatterings will
proceed in an analogous way.



\begin{thebibliography}{100}

\bibitem{EGS} Ehlers, J., Geren, P., \& Sachs, R. K., {\it
  J. Math. Phys.} {\bf 9} 1344 (1968).

\bibitem{FerMorPor99} Ferrando, J. J., Morales, J. A., \& Portilla, M., {\it Phys. Rev. D} {\bf 46} 578 (1992).

\bibitem{Clarkson:1999yj} Clarkson, C. \& Barrett, R., {\it
  Class. Quant. Grav.} {\bf 16} 3781 (1999).

\bibitem{2010CQGra..27l4008C} Clarkson, C. \& Maartens, R., {\it
  Class. Quant. Grav.} {\bf 27} 124008 (2010).

\bibitem{1995PhRvD..51.5942M} Maartens, R., Ellis, G. F. R., \& Stoeger, W. R.,
  {\it Phys. Rev. D} {\bf 51} 5942 (1995).

\bibitem{1995ApJ...443....1S} Stoeger, W. R., Maartens, R., \& Ellis, G. F. R.,
  {\it Astrophys. J.} {\bf 443} 1 (1995).

\bibitem{1995PhRvD..51.1525M} Maartens, R., Ellis, G. F. R., \& Stoeger, W. R.,
  {\it Phys. Rev. D} {\bf 51} 1525 (1995).

\bibitem{Rasanen:2009mg} R\"{a}s\"{a}nen, S., {\it Phys.\ Rev.\  D} {\bf 79} 
  123522 (2009).


\bibitem{HP} Hasse, W. \& Perlick, V., {\it Class. Quantum Grav.}
  {\bf16} 2559 (1999).

\bibitem{Roy} Maartens, R., {\it Phil. Trans. R. Soc. A} {\bf 369},
  5115 (2011).  

\bibitem{MM} Maartens, R. \& Matravers, D. R., {\it
  Class. Quant. Grav.} {\bf 11} 2693 (1994).

\bibitem{Hellaby} Mustapha, N., Hellaby, C., \& Ellis, G. F. R., {\it
  Mon. Not. Roy. Astron. Soc.} {\bf 292} 817 (1997).

\bibitem{2008JCAP...09..016G} Garc{\'{\i}}a-Bellido, J. \& Haugb{\o}lle,
  T., {\it JCAP} {\bf 9} 16 (2008).

\bibitem{2010CQGra..27f5002G} Garfinkle, D., {\it
  Class. Quant. Grav.} {\bf 27} 065002 (2010).

\bibitem{2011PhRvL.107d1301Z} Zhang, P. \& Stebbins, A., {\it
  Phys. Rev. Lett.} {\bf 107} 041301 (2011).

\bibitem{2011CQGra..28p4005Z} Zibin, J. P. \& Moss, A., {\it
  Class. Quant. Grav.} {\bf 28} 164005 (2011).

\bibitem{2011arXiv1108.2222B} Bull, P., Clifton, T., \& Ferreira,
  P. G., {\it Phys. Rev. D} {\bf 85}, 024002 (2012).

\bibitem{2010JCAP...10..011Y} Yoo, C. M., Nakao, K-I., \& Sasaki, M.,
  {\it JCAP} {\bf 10} 11 (2010).

\bibitem{2011JCAP...02..013C} Clarkson, C. \& Regis, M., {\it JCAP}
  {\bf 2} 13 (2011).

\bibitem{Goodman} Goodman, J., {\it Phys. Rev. D} {\bf 52} 1821 (1995).

\bibitem{CaldwellStebbins} Caldwell, R. R. \& Stebbins, A., {\it
  Phys. Rev. Lett.} {\bf 100} 191302 (2008).

\bibitem{2011CQGra..28p4004M} Marra, V. \& Notari, A., {\it Class. Quant. Grav.} {\bf 28} 164004 (2011).

\bibitem{tSZ} Sunyaev, R. A. \& Zel'dovich, Ya. B.,
  {\it Astrophys. Sp. Sci.} {\bf 7} 3 (1970).

\bibitem{kSZ}  Sunyaev, R. A. \& Zel'dovich, Ya. B., 
  {\it Mon. Not. Roy. Astron. Soc.} {\bf 190} 413 (1980).

\bibitem{Ellis}  Ellis, G. F. R., in {\it General Relativity and
  Cosmology}, Proc. Int. School of Physics ``Enrico Fermi'' (Varenna),
  p. 104 (R. K. Sachs, Ed.), Academic Press, San Diego (1971).

\bibitem{bb1} Chluba, J. \& Sunyaev, R. A., {\it A. \& A.} {\bf 424},
  389 (2004).

\bibitem{Maartens:1998xg}
  Maartens, R., Gebbie, T., \& Ellis, G. F. R.,
  {\it Phys.\ Rev.\ D}  {\bf 59}, 083506 (1999).

\bibitem{1997icm..book.....K} Krasi\'{n}ski, A.,  
{\it Inhomogeneous Cosmological Models}, p. 333, Cambridge University
Press, Cambridge (1997).

\bibitem{Clarkson:2001qc}
  Clarkson, C. A., Coley, A. A., \& O'Neill, E. S. D.,
  {\it Phys.\ Rev.\  D} {\bf 64}, 063510 (2001).

\bibitem{extra1} Quercellini, C. {\it et al.},  arXiv: 1011.2646.

\bibitem{extra2} Itoh, N. {\it et al.}, {\it Mon. Not. Roy. Astron. Soc.} {\bf 327}, 567 (2001).

\bibitem{extra3} Dolgov, A. D. {\it et al.}, {\it Astrophys. J.} {\bf 554}, 74 (2001).


\end{thebibliography}
\end{document}